\documentstyle[prl,twocolumn,aps]{revtex}
\begin{document}
\author{
L.~Y.~Gorelik$^\ast$, V.~S.~Shumeiko$^\ast$, R.~I.~Shekhter, G.~Wendin,
 and M.~Jonson}

\address{
Department of Applied Physics, Chalmers University of Technology
and G\"oteborg University, S-412 96 G\"oteborg, Sweden. \\
$^\ast$and B. I. Verkin Institute for Low Temperature Physics and Engineering,
47 Lenin Avenue, 310164 Kharkov, Ukraine.
}

\title{\bf Microwave-Induced ``Somersault Effect" in Flow of Josephson
Current through a Quantum Constriction
}
\maketitle
\begin{abstract}
%
%
%
We consider the supercurrent flow through gated mesoscopic semiconductor
hetrostructures in which a two-dimensional normal constriction is confined
between superconducting electrodes. We show that for these structures the
Josephson current, carried by quantized electron modes, can be strongly
affected
by an electromagnetic field. Photon-assisted Landau-Zener transitions between
Andreev bound states in the constriction manifest themselves in one of two
ways: (i) If the phase difference between the superconducting elements is fixed
a series of reversals in the direction of supercurrent flow occurs (Somersault
effect). (ii) If instead the junction is current biased a series of voltage
spikes is seen. Both manifestations are a direct consequence of the discrete
energy spectrum of the mesoscopic junction. We discuss necessary conditions
for
the  described phenomena to be experimentally observable.
\end{abstract}
\pacs{PACS numbers: } 
74.50.+r, 73.40.Rw \maketitle \narrowtext

The Josephson coupling between two superconductors is associated with
charge transfer through the non-superconducting region that separates them. The
coupling is therefore determined in a crucial way by the nature of the
electron states in this region. Quantum mechanical
tunneling through an insulating barrier (SIS-junction) \cite{Josephson},
itinerant propagation
of free electrons through a normal region (SNS) \cite{Kulik}, and propagation
of plasma
waves (boson modes) through a 1D-channel of strongly correlated electrons
\cite{Trieste} are examples of charge transfer mechanisms
in different types of weak links.

The electron states carrying the supercurrent in the non-superconducting
region can be influenced by an external time-dependent field. This possibility
raises interesting questions on the nature of Josephson coupling due to
nonequilibrium electron states. This is the problem addressed here.

Nonequilibrium effects can be expected to be particularily important in
situations where the Josephson coupling is mediated by a normally conducting
microconstriction, where only a few electron states --- Andreev bound states
---
 carry the current. Weak links through ballistic
microconstrictions in the two-dimensional (2D) electron gas of gated
semiconductor heterostructures have recently been observed experimentally
\cite{2DEG}.
An important feature of this structure is that the electron energy levels in
the constriction can be easily influenced by applying a potential to
the gate electrodes. In this way the electron concentration (and hence the
Fermi wavevector, $k_F$) in the channel can be controlled. The potential due to
induced charges on the gate electrodes furthermore provides a mechanism for
coupling the 2D electrons to an external electromagnetic field.

The geometry of the system to be considered in the following is shown in
Fig.~1. The Josephson coupling between two superconductors
can be expressed in terms of the phase difference $\phi$ between their
respective order parameters. We will first consider this phase difference to
be a
fixed quantitity, which in a SQUID geometry can be controlled entirely by the
magnetic flux (see  inset of Fig. 1). Andreev reflection at the boundaries
between superconducting and normal regions in a SNS configuration is known
\cite{Andreev} to lead to  a discrete set of energy levels
within the energy gap of the superconductors.
Due to spatial quantization
in the transverse $y$-direction, the energy spectrum of the Andreev bound
states
in a narrow and short constriction  consists  of a discrete set of pairs of
levels  labelled by the quantum number $n$ \cite{Be2},
\begin{equation}
 E_{n,\pm}= \pm \Delta \sqrt{1-T_n({k_F})\sin^2(\varphi/2)}.
\label{En12}
\end{equation}
Energy is measured from the Fermi level and
the transmission eigenvalue $T_n({k_F})$ is
related to the propagation of normal Fermi
level electrons through the junction (along the $x$-direction).  For a short
junction all of the Josephson current is carried by these Andreev states. It is
important to note that states below ($E_{n,-}$) and above ($E_{n,+}$) the
Fermi level carry current in different directions. In Fig.~2a, the position
of a pair of levels $E_{n,\pm}$ is indicated by the direction of the
corresponding partial currents marked by arrows. The dashed arrow for the
state $E_{n,+}$ above the Fermi energy illustrates that this state is
unoccupied at zero temperature and hence does not contribute to the total
current.

We shall consider a situation where a gate-induced time-dependent potential is
a sum of one part $V_0(t)$, which varies slowly on a scale
related to the interlevel spacing, and another rapidly oscillating part
$V_\omega\cos(\omega t)$, where $\omega$ is of the order of the interlevel
spacing. The high-frequency component of the induced potential produces
interlevel transitions, which result in a strong coupling of pairs of Andreev
states at resonance. Due to the change in the interlevel distances
resulting from a slow time variation of $k_F$
the condition for resonance,
$\omega = \omega_{\alpha\beta}\equiv (E_\alpha-E_\beta)/\hbar$ ($\alpha=
n,\pm$),
can be met for any pair of levels at some time $t=t_{{\rm res}}$ even if they
are
out of  resonance with the electromagnetic field at  $t=0$,
when the potential is switched on. Such a drift
into and out of resonance is illustrated in Fig.~2, where panel (a) shows a
non-resonant configuration. When the two levels drift into resonance, as in
Fig.~2b,  resonant interlevel mixing results in a finite population of the
upper level and a corresponding contribution to the partial Josephson current
in the reverse direction.

If the Andreev levels
pass slowly enough (see below) through the resonance, the result of the
dynamic evolution is a complete depopulation of the lower level. Hence, as
the levels drift out of resonance the Josephson current has been turned
around and the current is flowing in the opposite direction (Fig.~2c).
After some time out of resonance, the system relaxes back to its
ground state  and the Josephson current returns to the forward
direction. The  microwave-induced ``Somersault" has been completed.

To discuss the above effects quantitatively we begin with the time
dependent Bogoliubov-de Gennes (BdG) equation:
\begin{equation}
   i\hbar \dot{\bf \Psi}({\bf r},t)=
   \left(\hat H +\hat V(t) \right)
 {\bf \Psi}({\bf r},t).
   \label{BdG}
\end{equation}
Here $\hat H$ is the standard Hamiltonian for both the normal electrons in the
microconstriction part of the device, $|x|<L/2$, (see Fig.1)
and for the electrons in the superconducting regions, $|x|>L/2$.
The length of the junction $L$ is assumed to be small in comparison with
the coherence length $\xi_{Sm}=v_F/\Delta$ ($v_F$ is the Fermi velocity
in the constriction). The normal electron Hamiltonian
includes an electrostatic potential that confines the electrons
in the transverse direction and also describes the normal
scattering from impurities and from the  barriers at the Sm-S boundaries.
The spatial variation of the superconducting order parameter due to the
proximity
effect is small and can be neglected since the width of the constriction is
much
smaller than the superconductor coherence length $\xi_S$
\cite{KO}. Hence we use a step-function model for the pair potential,
$\Delta({\bf r})=\Delta\Theta(|x|-L/2)\exp\left(i\  {\rm
sign}(x)\phi/2\right)$.
The time dependent potential, $\hat V(t)=[V_0(t)+V_\omega\cos\omega
t]\sigma_z\Theta(L/2-|x|)$, where $\sigma_z$ is a Pauli spin matrix, is
confined
to the normal region and does not mix the electron- and hole-like components of
the wave function ${\bf \Psi}$. If its high frequency part $V_\omega$ is
absent,
the only effect of the external time-dependent field $V(t)$ is an adiabatic
energy shift of the Andreev levels caused by the parametric dependence of
$E_\alpha$ on $k_F=[(2m/\hbar^2)(\mu - V_0(t)]^{1/2}$ (see Eq. (\ref{En12})).
If
the field has a finite high frequency component, different Andreev states are
mixed due to interlevel transitions described by the matrix element  $
V_{\alpha\beta} = V_\omega \langle\Psi_\alpha({\bf r}),  \sigma_z
\Theta(L/2-|x|)\Psi_\beta({\bf r})\rangle. $
In what follows we will consider the weak coupling limit
$V_{\alpha\beta}\ll\hbar\omega_{\alpha\beta}$, in which almost all Andreev
states are only slightly renormalized by the electromagnetic field.
The exception is when the condition for resonance between two states,
$\omega = \omega_{\alpha\beta}(t)$, is
fullfilled at some time $t=t_{{\rm res}}$. At resonance the interaction
with the field cannot be treated perturbatively. Rather one must make
the resonant approximation by finding a solution to the BdG equation
(\ref{BdG}) that is a mixture of the two adiabatic states,
$\Psi_\alpha({\bf r})$ and $\Psi_\beta({\bf r})$, in
resonance:
\begin{eqnarray}
\Psi_{\alpha\beta}({\bf r},t) &=& \exp\left(-i/2\hbar\int^t
[E_\alpha(t)+E_\beta(t)]dt\right) \nonumber \\ && \times \left[ b_s^1
\Psi_\alpha({\bf r})e^{i(\omega/2)t} +
b_s^2\Psi_\beta({\bf r})e^{-i(\omega/2)t} \right].
\label{four}
\end{eqnarray}
Here the index $s=1,2$ labels independent solutions corresponding to
different initial conditions, $b_s^r(t=0)=\delta_{s,r}$.
After substituting the wave function (\ref{four}) into the BdG equation
(\ref{BdG}) and averaging over the fast oscillations, we readily find
equations which describe the time evolution of the coefficients
$b_s^{1,2}$. Introducing a vector coefficient ${\bf b}_s(t)$
we get
\begin{equation}
i \dot{\bf b}_s(t) = \left[ {\delta\omega_{\alpha\beta}(t)\over 2}\sigma_z
+ \left( {V_{\alpha\beta}\over\hbar}\sigma_+ +{\rm h.c.} \right)\right]
{\bf b}_s(t) ,
\label{six}
\end{equation}
where
$\delta\omega_{\alpha\beta}(t) =\omega_{\alpha\beta}(t)-\omega\ll\omega$
and $\sigma_+=(\sigma_x+i\sigma_y)/2$, $\sigma_{x,y}$ being
Pauli spin matrices. A problem similar to Eq.~(\ref{six}) was first discussed
by
Landau and Zener in connection with molecular pre-dissociation
\cite{Landau,Zener}. It follows from their theory that the asymptotic solution
of
(\ref{six}) valid at  $t\gg t_{{\rm res}}$ is
$
\left| b_s^r\right| ^2 = W_{\alpha\beta} +
\delta_{rs}\left(1-2W_{\alpha\beta}\right).
$
The probability $W_{\alpha\beta}$ for the system to be in a different
state after passing through the resonance,
\begin{equation}
W_{\alpha\beta}=1-{\rm e}^{-2\pi\gamma^2_{\alpha\beta}},
\end{equation}
is determined by the product of the interlevel transition
frequency  $|V_{\alpha\beta}|/\hbar$ and the characteristic time $\delta
t_{{\rm res}} = |\dot\omega_{\alpha\beta}(t_{{\rm res}})|^{-1/2}$ the system
spends on resonance,
\begin{equation}
\gamma_{\alpha\beta} = \left|V_{\alpha\beta}/\hbar\right|
\left|\dot\omega_{\alpha\beta}(t_{{\rm res}})\right|^{-1/2}.
\label{seven}
\end{equation}
If the characteristic time is long,
$\gamma\gg 1$ ($\delta t_{{\rm res}}\gg \hbar/V_{\alpha\beta}$), we have
$|b_s^r|^2\simeq (1-\delta_{r,s})$ and the  transition between modes $\alpha$
and
$\beta$ is complete.

In the absence of relaxation, the Josephson current is solely determined
by the dynamic evolution of the electron-hole wave function $\Psi({\bf
r},t)$ discussed above.
Before the high frequency field is switched on, the Andreev
levels are in equilibrium and the Josephson current has its equilibrium
value $I_0$. The deviation of the time averaged current, $\delta\bar I=\bar
I-I_0$, is found using the asymptotic result for the coefficients
$b_s^r(t)$:
\begin{equation}
\delta\bar I=\sum_{\{\alpha,\beta\}_{{\rm res}}}
W_{\alpha\beta}\left(n_\alpha-n_\beta \right)
\case{1}{2}\left(I_\beta-
I_\alpha\right)\left(2-\delta_{E_\alpha,-E_\beta}\right),
\label{ten}
\end{equation}
where $I_{\alpha,\beta}=(2e/\hbar)dE_{\alpha,\beta}/d\phi$, and
$n_{\alpha,\beta}$ is the Fermi function,
$n_{\alpha,\beta}=n_F(E_{\alpha,\beta})$.
The summation in (\ref{ten}) is over pairs of
levels, $\{\alpha,\beta\}_{{\rm res}}$, which have passed through a resonance
\cite{note}.  As we can see from Eq.~(\ref{ten}) only transitions between
states with energies of opposite signs (the Fermi level is at zero) contribute
to $\delta\bar I$. If the Josephson current through the channel is carried by a
single pair of Andreev levels, the direction of the Josephson current is
reversed for $W\geq 1/2$. It is essential to note that for $t\gg t_{{\rm res}}$
there is no resonant coupling with the electromagnetic field to
maintain the inverse population of Andreev levels responsible for the reversal
of the current. Any inelastic relaxation mechanism will bring the system back
to
its equilibrium state and restore the original direction of Josephson
current. This process of dynamically switching on the inverse
population of Andreev states and the subsequent inelastic relaxation
manifests itself as a `Somersault' of the Josephson current.

Equation (\ref{ten}) is significantly simplified if
the microconstriction joining the two superconductors is an adiabatically
smooth ballistic channel with quantized energy levels (modes) corresponding to
the transverse electron motion. In this situation each propagating mode
produces
one pair of Andreev levels of the type (\ref{En12}) --- one level above- and
one
below the Fermi  energy --- and the transmission eigenvalue $T_n$ reduces to
the
normal electron transmission coefficient of the $n$-th mode,
\begin{equation}
T_n(k_F)=\left[1+\frac{4R}{(1-R)^2}\sin^2(k_n L)\right]^{-1},\\
\end{equation}
where $k_n=\sqrt{k_F^2 - \pi^2 n^2/d^2}$, and
$R$ is the probability for electron backscattering at the Sm-S boundary.
The electromagnetic field couples levels only within these pairs, i.e. if
$\sigma=\pm$ one has
$V_{\alpha\alpha^\prime}=\delta_{n,n^\prime}
\delta_{\sigma,-\sigma^\prime}V_n$
and consequently
$\omega_{\alpha\alpha^\prime}=\delta_{n,n^\prime}
\delta_{\sigma,-\sigma^\prime}\omega_n$.
The matrix element $V_n$ has the form:
\begin{equation}
|V_n|^2 = V_\omega^2 \frac{L^2}{\xi_{{\rm Sm}}^2} T^3_n\frac{R}{(1-R)^2}
\frac{\cos^2(k_n L)\sin^2\phi}{1-T_n\sin^2(\phi/2)}.
\label{twelve}
\end{equation}
In the ballistic case the total current at zero temperature, taking into
account Eq.~(\ref{ten}), can be written in the form:
\begin{equation}
\bar I = \sum_{n} I_n (1-2W_n),
\label {thirteen}
\end{equation}
where the summation is formally done over all transverse modes and $W_n$ is
assumed to be zero for those Andreev levels which
have not passed through a resonance.

As one can see from Eq.~(\ref{twelve}), the electromagnetic field can
only mix Andreev states if the probability for electron backscattering at the
Sm-S boundaries is finite. A `Somersault'  which changes
the total momentum of the system can evidently only occur if backscattering is
possible.
It is interesting that if there is a geometric resonance, $k_n
L=n\pi$, the Somersault effect still takes place according to (\ref{ten})
even though the microconstriction is fully transparent for electrons in mode
$n$, $T_n(k_F)=1$. This fact provides an important possibility of
distinguishing
between resonant transmission through the double barrier structure and
balllistic propagation in the absence of normal backscattering from Sm-S
boundaries.

It is also interesting to consider the case when there is no drift with time
of the Andreev levels, i.e. when $V_0(t)=0$. A permanent resonant
coupling of Andreev states is now possible and leads to an equalization of
the average level population \cite{Sh}. Solving
equations (\ref{six}) we find a result for the Josephson
current that deviates from its stationary value:
\begin{equation}
\bar I = \sum_n I_n \frac{|\delta\omega_{n}|}
{\sqrt{\delta\omega_{n}^2 +
\left(2 V_{n}/\hbar\right)^2}}.
\label{fourteen}
\end{equation}
In particular we find that resonant coupling, i.e.
when $\delta\omega_{n}=0$, results in a complete blockade of the
Josephson current. This phenomenon can be observed as a sequence of dips if
the Josephson current is plotted as a function of the frequency of the
electromagnetic field. The number of dips should equal the number of
propagating modes in the ballistic constriction and the dip amplitude should
equal the single mode contribution $I_n$ to the stationary Josephson
current.

In the discussion so far we have assumed that the phase difference
between the two superconductors is constant in time. For the Somersault
effect to develop as described above it is necessary that $\phi$ be fixed to a
high degree of accuracy, otherwise the system will escape from resonance. This
fact implies that the change of the interlevel distance due to the phase shift
$\delta\phi$ should be smaller than the nonlinear resonance width:
$
\hbar(\partial\omega_{\alpha\beta}/\partial\phi) \delta\phi\ll
V_{\alpha\beta}
$.
To estimate $\delta\phi$ we first observe that $\phi$
is related to the shielding currents induced by the enclosed magnetic flux in
the SQUID geometry. This Meissner current flows
near the surfaces in the thick part of the ring shown in Fig.~1. The diameter
of
the ring cross section is $d\gg \lambda$, where $\lambda$ is the
penetration length of the magnetic field. The current flowing through the
small ring segment that forms a narrow channel is much smaller than the
Meissner
current in the main part of the ring. The Somersault effect, which affects
this small part of the induced current, will result in a redistribution of
Meissner current in the bulk ring region. The corresponding phase shift will
be of order $\delta\phi \sim ({\cal L}/\xi_S)(1/N_\lambda)$, where ${\cal L}$
is the ring length, $\xi_S$ is the superconductor coherence length, and
$N_\lambda\sim k_0^2\lambda d$ is the number of transverse modes carrying the
Meissner current (here $k_0$ is the Fermi wave vector in the superconducting
ring).  The criterion for the phase difference to be stable enough to observe
the resonance is thus $(V_{\alpha\beta}/\Delta)(L/{\cal L}) \gg
(1/N_\lambda)$.

When the Josephson junction is connected to a current source, the current
rather
than the phase difference is held constant. In this situation another
interesting phenomenon occurs; the Somersault effect causes phase
slips at times when Andreev levels are resonantly coupled. The phase slips in
turn manifest themselves as voltage spikes. The
characteristic time scale of a phase slip is determined by the Josephson
plasma frequency $\omega_J=[(\Delta/\hbar)(1/R_0C_J)]^{1/2}$, which for the
geometry of Fig.~1 and $R_0=\hbar/e^2$ can be written as
$\omega_J=(\Delta/\hbar)(\xi_S L/d^2)^{1/2}$. This relationship allows us to
estimate the amplitude and the duration of the voltage spike as
$U=(\hbar/2e)\dot\phi\sim(\hbar/e)\omega_J$  and  $\Delta t\sim 1/\omega_J$
respectively in the case where the duration of the
interlevel transition, $\delta t_{{\rm res}}$, is shorter than $\Delta t$ i.e.
if
$1/\omega_J \gg \delta t_{{\rm res}} $. In the opposite case,
$1/\omega_J \ll \delta t_{{\rm res}} $, a
more complex analysis of the joint dynamics of the phase- and interlevel
transition gives the estimates
$U\sim
V_{\alpha\beta}^{4/3}/e\Delta^{1/3}$, and
$\;\;\Delta t\sim
\hbar/V_{\alpha\beta}^{2/3}\Delta^{1/3}$.

The Somersault effect discussed above relies on the preservation of phase
coherence during the dynamical evolution of the Andreev states and the
redistribution of level populations at resonance.
This implies that the phase breaking time $\tau$ associated with inelastic
relaxation should be the largest time scale in the problem:
\begin{equation}
1/\omega_{\alpha\beta}\sim \hbar/\Delta \ll \hbar/V_{\alpha\beta}\ll
\delta t_{{\rm res}} \ll \tau,
\;\;\;1/\omega_J\ll \tau.
\label{eighteen}
\end {equation}
Since the main part of the Andreev state wave function is localized
in the superconducting electrodes the most important
relaxation mechanisms are recombination processes due to
electron-phonon interaction in the superconductors. According to
Ref.~\cite{Kaplan} a typical estimate of the recombination time in a
superconductor is  $\Delta\tau/\hbar>$ 10$^2$. This value makes it possible
to fullfil the inequalities (\ref{eighteen}) in a device with normal region
length $L\approx\xi_{Sm}$ if the
amplitude $V_\omega$ is of order of 10 $\mu$V. Meanwhile, an amplitude of the
voltage spikes in a current biased junction is of order of 1-10 $\mu$V.

In conclusion, we have shown that a microwave electromagnetic field can
drastically change the Josephson current carried by quantized modes in
a mesoscopic constriction, giving rise to sequential reversals of the
current in a phase biased junction and to voltage spikes in a current
biased junction. Observation of the effect, which seems to be possible
in currently available gated semiconductor heterostructure devices, would be a
direct manifestation of the discrete nature of the Josephson current in a
quantum
constriction.

This work was supported by the Swedish Institute, the Swedish Natural
Science Research Council, and the Swedish Research Council for
Engineering Sciences.

\begin{figure}
\caption{
Sketch of a gated semiconductor heterostructure in which a two-dimensional
normal constriction is confined between superconducting electrodes. Inset:
The phase difference between the two
superconductors can be kept constant in the SQUID configuration shown
(see text).
}
\end{figure}

\begin{figure}
\caption{ Microwave induced transitions and currents carried by the Andreev
levels in the normal constriction of Fig.~1 during a slow drift of the levels
through resonance:  (a) before reaching
the resonance; (b) during the resonance; (c) after passing the resonance.
Arrows show the direction of current, dashed arrow correspond to unpopulated
level.}
\end{figure}


\begin{references}
\bibitem{Josephson} B.D. Josephson, Phys. Lett. {\bf 1}, 251 (1962).
\bibitem{Kulik} I.O. Kulik, Sov. Phys. JETP {\bf 30}, 944 (1970).
\bibitem{Trieste} R. Fazio, F.W.J. Hekking, and A.A. Odintsov, Preprint
(1994).
\bibitem{2DEG} A. Dimoulas, J.P. Heida, B.J. van Wees, T.M. Klapwijk, W.
van der Graaf, and G. Borghs, Phys. Rev. Lett., {\bf 74}, 602 (1994).
\bibitem{Andreev} A.F. Andreev, Sov. Phys. JETP {\bf 19}, 1228 (1964).
\bibitem{Be2} C.W.J.Beenakker, Phys. Rev. Lett. {\bf 67}, 3836 (1991).
\bibitem{KO} I.O. Kulik and A.N. Omel'yanchuk, Sov. J. Low Temp. Phys., {\bf
3}, 459 (1977).
\bibitem{note} Eq.(\ref{ten}) is derived under the assumption that a state
passes
resonance not more than once during a typical recombination time. Otherwise the
current structure is quite intricate due to interference effects of multiple
resonance transitions.
\bibitem{Landau} L.D. Landau, Phys. Z. Sowijet {\bf 1}, 88 (1932).
\bibitem{Zener} C. Zener, Proc. R. Soc., Ser.A {\bf 137}, 696 (1932);
{\bf 140}, 660 (1933).
\bibitem{Sh} V.S. Shumeiko, G. Wendin, and E.N. Bratus', Phys. Rev. B {\bf
48}, 13129 (1993).
\bibitem{Kaplan} S.B. Kaplan, C.C. Chi, D.N. Langenberg, J.J. Chang, S.
Jafaney, and D.J. Scalapino, Phys. Rev. B {\bf 14}, 4854 (1976).

\end{references}
\end{document}